\documentclass[useAMS,usenatbib]{mn2e}

\usepackage{amssymb}
\usepackage{times}
\usepackage{graphicx}
\usepackage{subfigure}

\newcommand{\beq}{\begin{equation}}
\newcommand{\eeq}{\end{equation}}

\newcommand{\bfd}{\mathbf{d}}

\newcommand{\bfP}{\mathbf{P}}

\def\gs{\mathrel{\lower0.6ex\hbox{$\buildrel {\textstyle >}\over{\scriptstyle \sim}$}}}
\def\ls{\mathrel{\lower0.6ex\hbox{$\buildrel {\textstyle <}\over{\scriptstyle \sim}$}}}
\newcommand{\simgt}{\lower.5ex\hbox{$\; \buildrel > \over \sim \;$}}
\newcommand{\simlt}{\lower.5ex\hbox{$\; \buildrel < \over \sim \;$}}

\newcommand{\aap}{A\&A}
\newcommand{\apj}{ApJ}
\newcommand{\apjl}{ApJ}
\newcommand{\apjs}{ApJS}

\newcommand{\mnras}{MNRAS}
\newcommand{\physrep}{Phisycs Rep.}
\newcommand{\apss}{Astroph. Sp. Science}

\begin{document}

\title[A1689 shape and orientation]{Shape and orientation of the gas distribution in A1689}
\author[M. Sereno, S. Ettori and A. Baldi]{
M. Sereno$^{1,2}$\thanks{E-mail: mauro.sereno@polito.it (MS)}, S. Ettori$^{3,4}$ and A. Baldi$^{5,3}$
\\
$^1$Dipartimento di Fisica, Politecnico di Torino, corso Duca degli Abruzzi 24, I-10129 Torino, Italia\\
$^2$INFN, Sezione di Torino, via Pietro Giuria 1, I-10125, Torino, Italia\\
$^3$INAF, Osservatorio Astronomico di Bologna, via Ranzani 1, I-40127 Bologna, Italia\\
$^4$INFN, Sezione di Bologna, viale Berti Pichat 6/2, I-40127 Bologna, Italia\\
$^5$Dipartimento di Astronomia, Universit\`a di Bologna, via Ranzani 1, IÐ40127, Bologna, Italy
}


\maketitle

\begin{abstract}
Knowledge of intrinsic shape and orientation of galaxy clusters is crucial to understand their formation and evolution. We propose a novel model which uses Bayesian inference to determine the intrinsic form of the hot intracluster medium of galaxy clusters. The method exploits X-ray spectroscopic and photometric data plus measurements of the Sunyaev-Zel'dovich effect (SZe). The gas distribution is modelled with an ellipsoidal parametric profile who can fit observed X-ray surface-brightness and temperature. Comparison with the SZ amplitude fixes the elongation along the line of sight. Finally, Bayesian inference allows us to deproject the measured elongation and the projected ellipticity and constrain the intrinsic shape and orientation of the cluster. We apply the method to the rich cluster Abell 1689, which was targeted by the \textit{Chandra} and {\it XMM} satellites as well as by several SZe observatories. Observations cover in detail a region $\ls 1~$Mpc. Our analysis favours a mildly triaxial cluster with a minor to major axis ratio of $0.70 \pm 0.15$, preferentially elongated along the line of sight, as expected for massive lensing clusters. The triaxial structure together with the orientation bias can reconcile X-ray with lensing analyses and supports the view of A1689 as a just slightly over-concentrated massive cluster not so far from hydrostatic equilibrium. 
\end{abstract}


\section{Introduction}

Clusters of galaxies are the most recent bound structures to form in the Universe \citep{voi05}. Their intrinsic shape contains evidence of the assembly process of structures. Accurate knowledge of forms is important on its own by probing the cosmic structure formation. It suggests how material aggregates from large-scale perturbations \citep{wes94,ji+su02} and contains evidence about the nature and mechanisms of interaction of baryons and dark matter \citep{lee+sut03,kaz+al04}. 

The complex structure of halos also affects the estimation of quantities which are crucial in any attempt at high precision cosmology. Favouring a simple spherical model over more realistic shapes might cause a significant bias in estimating the cluster mass \citep{gav05}, the inner matter density slope and the concentration \citep{ogu+al05}.  

Accurate knowledge of the intrinsic structure is also critical when comparing observations with theoretical predictions. The observed concentration-mass relation for galaxy clusters has a slope consistent with what found in $N$-body numerical simulations, though the normalization factor is higher \citep{co+na07,ett+al10}. Disagreement between theory and observation might be explained by orientation and shape biases. In fact, triaxial halos can be much more efficient lenses than their more spherical counterparts \citep{og+bl09} with the strongest lenses in the Universe expected to be a highly biased population preferentially orientated along the line of sight. 

Classical attempts to determine intrinsic three dimensional forms were based on statistical approaches consisting in the inversion of the distribution of apparent shapes \citep{hub26,noe79,bin80,bi+de81,fa+vi91,det+al95,moh+al95,bas+al00,coo00,th+ch01,al+ry02,ryd96,pli+al04,paz+al06}. With the exception of disc galaxies, either prolate-like or triaxial shapes appear to dominate all cosmic structure on a large scale. \citet{kaw10} found that the observed probability density function of the projected axis ratio of a sample of X-ray clusters was compatible with that of a population of triaxial halos in hydrostatic equilibrium consistent with $N$-body simulations.

The determination of the form of single objects is a more recent topic \citep{zar+al98,reb00,dor+al01,pu+ba06}. Clusters of galaxy are very interesting targets since they can be probed with very heterogeneous data-sets at very different wave-lengths from X-ray surface brightness and spectral observations of the intra-cluster medium (ICM), to gravitational lensing (GL) observations of the total mass distribution to the Sunyaev-Zel'dovich effect (SZe) in the radio-band. On the theoretical side, it has been ascertained that the deprojection is not unique \citep{ryb87,ge+bi96}. Even assuming the cluster to be a triaxial ellipsoid, the only quantity that can be univocally determined is the elongation along the line of sight \citep{ser07}. This allows to break the degeneracy with the distance and to get an unbiased estimate of the Hubble constant \citep{fo+pe02}.

On the observational side, only a few works have tried to infer shape or orientation of single objects. Combined use of X-ray and SZe data allows to constrain the shape of the ICM without any assumption regarding equilibrium or geometry. \citet{def+al05} and \citet{ser+al06} performed a parametric analysis of a sample of 25 clusters, finding that prolate rather than oblate shapes seem to be preferred, with signs of a more general triaxial morphology. \citet{ma+ch11} derived a model-independent expression for the minimum line-of-sight extent of the hot plasma and applied it to the Bullet Cluster. A different approach is based on lensing observations. Surface density maps of the total matter distribution from weak or strong lensing have been deprojected exploiting some a priori assumptions on their intrinsic shapes \citet{ogu+al05,cor+al09,ser+al10,ser+al10b,se+um11}.

Here, we propose a parametric method to infer the intrinsic shape and orientation of clusters based on deep X-ray and SZe observations. First, we model the gas density and fit to the data to obtain a direct measurement of the elongation along the line of sight. Then, we exploit Bayesian methods which enable to make an inference about a number of variables larger than the number of the observed data. This allows to constrain both intrinsic shape and orientation.

\begin{figure}
       \resizebox{\hsize}{!}{\includegraphics{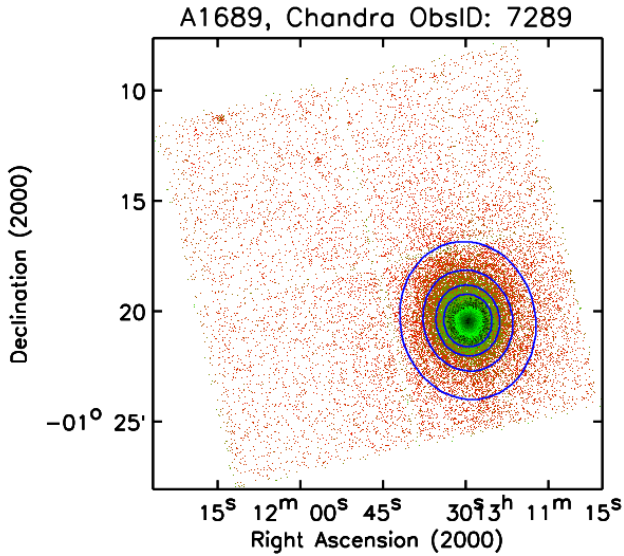}}
       \caption{Exposure-corrected image of one of the \textit{Chandra} observation used for the spatial analysis. The full lines are the best-fit ellipses enclosing 50, 60, 70 and 80 per cent of the light for a fixed X-ray centroid.}
	\label{a1689_ell}
\end{figure}

The method is applied to Abell 1689 (A1689), see Fig.~\ref{a1689_ell}, a very luminous cluster at redshift $z=0.183$ \citep{bro+al05,lim+al07}. Strong lensing analyses have provided a consistent picture of the mass distribution in the inner $\ls 300~\mathrm{kpc}$ regions \citep{bro+al05,hal+al06,lim+al07,coe+al10} favouring a quite concentrated mass distribution. On the other hand, some disagreement persists on the larger virial scale. Different weak lensing analyses suggest somewhat different degrees of concentration \citep{ume+bro08,ume+al09,cor+al09,lim+al07}. There is some conflict between X-ray and lensing analyses, with lensing masses exceeding estimates derived under the hypothesis of hydrostatic equilibrium by 30-40 per cent in the inner regions. \citet{lem+al08} combined \textit{Chandra} X-ray brightness measurements and joint strong/weak lensing measurements under the hypothesis of hydrostatic equilibrium. They found that the resulting equilibrium temperature exceeds the observed temperature by 30 per cent at all radii.

Disagreement can be reduced properly accounting for deviations from either spherical symmetry or hydrostatic equilibrium. The level of hydrostatic equilibrium in A1689 has been recently reassessed. \citet{kaw+al10} found regions with low gas temperatures and entropies deviating from hydrostatic equilibrium. \citet{mol+al10} showed that in the core region of relaxed clusters alike A1689 a significant non-thermal pressure support of $\sim 20$ per cent is originated by subsonic random gas motions within 1--10 per cent of the virial radius.

A number of triaxial lensing analyses of A1689 has been performed \citep{ogu+al05,cor+al09}. Recently, \citet{se+um11} developed a method for a full three-dimensional analysis of strong and weak lensing data. They found evidence for a mildly triaxial lens (minor to major axis ratio $\sim 0.5 \pm 0.2$) with the major axis orientated along the line of sight. The triaxial shape is compatible with a halo slightly over-concentrated but still consistent with theoretical predictions. \citet{pen+al09} suggested that a prolate ellipsoidal configuration for the gas distribution, aligned with the line of sight and with an axis ratio of $\sim 0.6$, could solve the central mass discrepancy between lensing and X-ray estimates. Recently, \citet{mor+al11} combined lensing and X-ray data under some very restrictive hypotheses. They assumed the cluster to be aligned with the line of sight and shaped deviations from hydrostatic equilibrium with a very peculiar modelling. They also exploited results from numerical simulations relating the shapes of gas and matter distribution, which are affected by quite large errors for very elongated clusters \citep{lee+sut03}. They found an axial ratio for the matter distribution of $\simeq 0.5$.

Here, we address the shape shape determination of A1689 with a novel model. The paper is organised as follows. In Sec.~\ref{sec_tria}, we discuss the relation between intrinsic and projected quantities of triaxial ellipsoids. Section~\ref{sec_proj} lists some measurable X-ray and SZe quantities. Observations and data reduction are described in Sec.~\ref{sec_data}. In Sec.~\ref{sec_dist} we detail how the gas distribution was modelled and fitted to the data. Section~\ref{sec_addi} lists some sources of systematic errors. In Sec.~\ref{sec_depr} we perform the deprojection. Section~\ref{sec_conc} is devoted to some final considerations. Throughout the paper, we assume a flat $\Lambda$CDM cosmology with density parameters $\Omega_\mathrm{M}=0.275$, $\Omega_{\Lambda}=0.725$ and Hubble constant $H_0=100h~\mathrm{km~s}^{-1}\mathrm{Mpc}^{-1}$, $h=0.702 \pm 0.014$ \citep{kom+al11}. At the A1689 distance, $1\arcsec$ corresponds to $3.08~\mathrm{kpc}$.

\section{Triaxial ellipsoids}
\label{sec_tria}

\begin{figure}
\resizebox{\hsize}{!}{\includegraphics{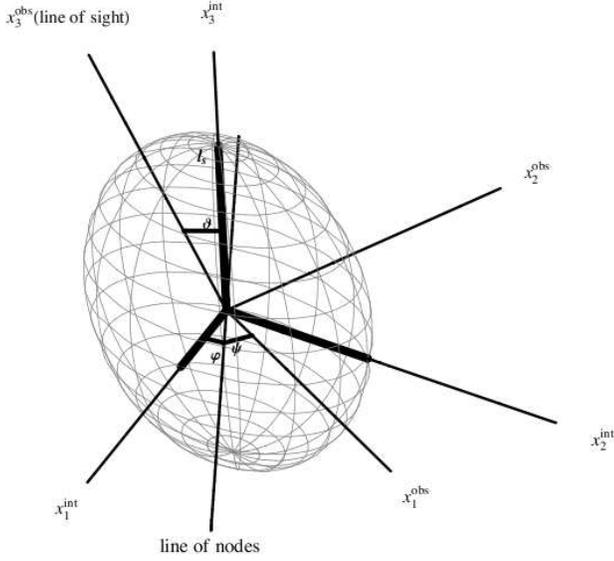}}
\caption{The orientation of a triaxial ellipsoid. The $x_i^\mathrm{int}$-coordinate axes are oriented along the principal axes of the ellipsoid and define the intrinsic reference system. The $x_i^\mathrm{obs}$-axes defines the observer's frame. The line of sight is oriented along the $x_3^\mathrm{obs}$-axis. The Euler's angle $\vartheta$, $\varphi$ and $\psi$ are displayed. Thick lines denote the ellipsoid axes. $l_\mathrm{s}$ is the major axis.
}
\label{fig_Euler_angles}
\end{figure}

\begin{figure}
\resizebox{\hsize}{!}{\includegraphics{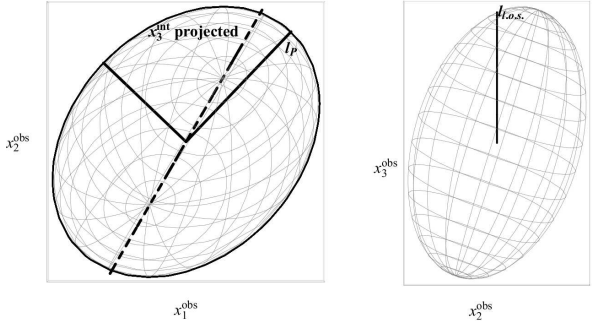}}
\caption{The ellipsoid of Fig.~\ref{fig_Euler_angles} as seen from the observer. \textit{Left panel}: projection of the ellipsoid into the plane of the sky. The observer sees an ellipse with major axis $l_\mathrm{p}$. The axes of the ellipse are plotted as thick lines. The dashed line is the projection in the plane of the sky of the $x_3^\mathrm{int}$-axis. \textit{Right panel}: ellipsoid as seen from above. This would be the view of an observer perpendicular to the line of sight, i.e., located on the $x_1^\mathrm{obs}$-axis. $l_\mathrm{l.o.s.}$ is the half size along the line of sight ($x_3^\mathrm{obs}$-axis).
}
\label{fig_ellipsoid_projection}
\end{figure}

High resolution $N$-body simulations have shown that the density profiles of massive halos are aspherical \citep{ji+su02,lee+sut03}. The electron density $n$ of the intra-cluster medium (ICM) can be assumed to be constant on a family of similar, concentric, coaxial ellipsoids. These assumptions were also observationally verified \citep{kaw10}.

The gas distribution in clusters of galaxies in hydrostatic equilibrium traces the gravitational potential. Since we are considering a triaxial elliptical gas distribution, the gravitational potential turns out to be ellipsoidal too. These gravitational potential can turn unphysical for extreme axial ratios, giving negative density regions or very unlikely configurations, but as far as either inner regions or small eccentricities are considered, they can provide very suitable approximations. 

An ellipsoidal ICM profile can be expressed as a function of only one radial variable $\zeta$, which runs along the major axis and replaces the spherical radius. The minor (intermediate) to major axial ratio is denoted as $q_1$ ($q_2$) with $0< q_1 \le q_2\le 1$; we also use the inverse ratios, $0< e_i =1/q_i \ge 1$. For a prolate shape, $q_1=q_2 \le 1$; a oblate ellipsoid has $q_1 \le q_2 =1$.

Three Euler's angles, $\vartheta, \varphi$ and $\psi$, relate the intrinsic to the observer's coordinate system, see Fig.~\ref{fig_Euler_angles}. The angle $\vartheta$ quantifies the inclination of the major axis with respect to the line of sight. $\vartheta$ and $\varphi$ fix the orientation of the line of sight in the intrinsic system. The third angle $\psi$ determines the orientation of the cluster in the plane of the sky. Since we are interested in the cluster shape, we can neglect this angle in the following.

When viewed from an arbitrary direction, quantities constant on similar ellipsoids project themselves on similar ellipses~\citep{sta77,ser07}, see Fig.~\ref{fig_ellipsoid_projection}. The ellipticity and the orientation of the projected ellipses depend only on the intrinsic geometry and orientation of the system. The axial ratio of the major to the minor axis of the observed projected isophotes, $e_{\rm p}(\geq 1)$, can be written as \citep{bin80},
\begin{equation}
\label{eq:tri4e}
e_{\rm p}= \sqrt{ \frac{j+l + \sqrt{(j-l)^2+4 k^2 } }{j+l -\sqrt{(j-l)^2+4 k^2 }} },
\end{equation}
where  $j, k$ and $l$ are defined as
\begin{eqnarray}
j & = &  e_1^2 e_2^2 \sin^2 \vartheta + e_1^2 \cos^2 \vartheta \cos^2 \varphi   +  e_2^2 \cos^2 \vartheta \sin^2 \varphi  ,  \label{eq:tri4a} \\
k & = &  (e_1^2 - e_2^2) \sin \varphi \cos \varphi  \cos \vartheta   ,  \label{eq:tri4b}  \\
l & = &  e_1^2 \sin^2 \varphi + e_2^2 \cos^2 \varphi . \label{eq:tri4c}
\end{eqnarray}
We also use the ellipticity $\epsilon = 1-1/e_\mathrm{P}$.

Ellipsoids map into ellipses. Let $l_\mathrm{s}$ be the intrinsic major axis of the ellipsoid, see Fig.~\ref{fig_Euler_angles}. The corresponding projected ellipse on the plane of the sky has a major axis $l_\mathrm{p}$, see Fig.~\ref{fig_ellipsoid_projection},
\begin{equation}
\label{eq:tri6}
\l_{\rm p} \equiv l_{\rm s} \left( \frac{e_{\rm p}}{e_1 e_2} \right)^{1/2} f^{1/4}
\end{equation}
where $f$ is a function of the cluster shape and orientation,
\begin{equation}
\label{eq:tri3}
f = e_1^2 \sin^2 \vartheta  \sin^2 \varphi  + e_2^2 \sin^2 \vartheta  \cos^2 \varphi + \cos^2 \vartheta  .
\end{equation}

The quantity 
\beq
l_\mathrm{l.o.s.}= l_\mathrm{s}/\sqrt{f} 
\eeq
is the half-size of the ellipsoid along the line of sight, i.e., as seen from above, see Fig.~\ref{fig_ellipsoid_projection}. Projection and elongation are related through \citep{ser07}
\beq
\label{mult1}
l_\mathrm{l.o.s.} \equiv \frac{l_\mathrm{P}}{e_\Delta},
\eeq 
which represents the definition of the elongation $e_\Delta$. In terms of intrinsic parameters,
\beq
e_\Delta  \equiv \left( \frac{e_\mathrm{P}}{e_1 e_2}\right)^{1/2} f^{3/4}.
\eeq
$e_\Delta$ is the ratio between the major axis of the projected ellipse in the plane of the sky and the size of the ellipsoid along the line of sight. It quantifies the elongation of the triaxial ellipsoid along the line of sight. If $e_\Delta < 1$, then the cluster is more elongated along the line of sight than wide in the plane of the sky, i.e., the smaller $e_\Delta$, the larger the elongation along the line of sight.

\section{Projection}
\label{sec_proj}

The projected map $F_\mathrm{2D}$ of a volume density $F_\mathrm{3D}$, which is constant on surfaces of constant ellipsodial radius $\zeta$, is elliptical on the plane of the sky \citep{sta77,ser07,ser+al10}. $F_\mathrm{2D}$  has the same functional form of a spherically symmetric halo,
\beq
\label{eq_proj1}
F_\mathrm{2D} (\xi; l_\mathrm{P}, p_i) = \frac{2 l_\mathrm{P}}{e_\Delta} \int_{x_\xi}^\infty F_\mathrm{3D}(x_\zeta; l_\mathrm{s}, p_i) \frac{x_\zeta}{\sqrt{ x_\zeta^2-x_\xi^2}} d x_\zeta,
\eeq
where $x_\zeta \equiv \zeta/l_\mathrm{s}$ is the integration variable, $\xi$ is the elliptical radius in the plane of the sky (measured along the projected major axis), $x_\xi \equiv \xi/l_\mathrm{P}$, $l_\mathrm{s}$ is the length scale of the 3D density, $l_\mathrm{P}$ is its projection on the plane of the sky, and $p_i$ are the other parameters describing the intrinsic density profile (inner or outer slope, tidal radius, ...). If we know the form of the intrinsic profile $F_\mathrm{3D}$, the only unknown in the projection is the undetermined geometrical factor $e_\Delta^{-1}$.

\subsection{X-ray Surface Brightness}

Cluster X-ray emission is due to bremsstrahlung and line radiation resulting from electron-ion collisions in the high temperature plasma ($k_{\rm B} T_\mathrm{e} \approx 8$-$10\ {\rm keV}$, with $k_{\rm B}$ being the Boltzmann constant). The observed X-ray surface brightness (SB) can be written as
\begin{equation}
SB^\mathrm{Obs} = \frac{1}{4 \pi (1+z )^3} \int _\mathrm{l.o.s.} n_\mathrm{e}^2 \Lambda_\mathrm{e}^\mathrm{Eff}(T_\mathrm{e}, {\cal{Z}}) dl ,
\label{eq:sxb1}
\end{equation}
where $T_\mathrm{e}$ is the intrinsic temperature, ${\cal{Z}}$ the metllicity and $\Lambda_\mathrm{e}^\mathrm{Eff}$ is the effective cooling function of the ICM in the cluster rest frame, which depends on the energy-dependent effective area of the instrument \citep{ree+al10}.  We computed predicted emissivities through the XSPEC software package \citep{arn96} adopting an optically-thin plasma (mekal) emission model .

\subsection{The Sunyaev-Zel'dovich effect}

Photons of the cosmic microwave background (CMB) that pass through the hot ICM of a cluster interact with its energetic electrons through inverse Compton scattering, slightly distorting the CMB spectrum. This is the Sunyaev-Zeldovich effect (SZe)~\citep{su+ze70,bir99}, which is proportional to the electron pressure integrated along the line of sight. The amplitude of the signal is proportional to the Compton-$y$ parameter,
\beq
\label{eq:sze1}
y \equiv \frac{ \sigma_{\rm T} k_{\rm B} }{m_\mathrm{e} c^2} \int_\mathrm{l.o.s.} n  T dl ,
\eeq
where $\sigma_{\rm T}$ is  the Thompson cross section, $m_\mathrm{e}$ the electron mass, $c$ the speed of light in vacuum.

The measured temperature decrement $\Delta T_\mathrm{SZ}$ of the CMB for an isothermal plasma is given by
\begin{equation}
\label{eq:sze2}
\Delta T_\mathrm{SZ} = f_\mathrm{SZ}(\nu, T) T_\mathrm{CMB} y
\end{equation}
where $T_\mathrm{CMB}$ is the temperature of the CMB and $f_\mathrm{SZ}(\nu, T)$ accounts for relativistic corrections at frequency $\nu$.

An overall measure of the thermal energy content in a cluster is given by $Y$, which is the Compton-$y$ parameter  integrated from its center out to a projected radius $r_\Delta$,
\beq
\label{eq:sze3}
Y_\Delta = \int_{\Omega_\Delta}y d \Omega
\eeq
where $\Omega_\Delta$ is the solid angle of the integrated patch covered within a radius $r_\Delta$. $Y$ is a more robust quantity for observational tests than the central value of Compton-$y$, because it is less dependent on the model of gas distribution used for the analysis \citep{ben+al04}. In addition, integrating the Compton-$y$ out to a large projected radius diminishes (though does not completely remove) effects resulting from the presence of strong entropy features in the central regions of clusters \citep{mca+al03}.

\subsection{Observed temperature}
The temperature $T_\mathrm{e}$ in Eqs.~\ref{eq:sxb1} and \ref{eq:sze1} is meant to be the intrinsic temperature. The spectroscopic temperature $T_\mathrm{P}$ measured by space observatories is well approximated by \citep{maz+al04}
\beq
\label{eq_proj7}
T_\mathrm{P}= \frac{\int W T_\mathrm{e} dV}{\int W dV}, \, \, W=\frac{n^2}{T_\mathrm{e}^{3/4}} .
\eeq

\section{Data}
\label{sec_data}

\subsection{X-ray surface brightness}

\begin{table}
\centering
\caption{Measured ellipticities from \textit{Chandra} data. The rotation angle $\theta_\epsilon$ is measured counterclockwise. The Position Angle PA ($=\theta_\epsilon-90\deg$) is defined equal to zero at North.}
\label{tab_ellip}
\begin{tabular}[c]{ccc}
        \hline
        \noalign{\smallskip}
	Exposure ID	& $\epsilon$ & $\theta_\epsilon$ \\ 
        \noalign{\smallskip}
         \hline
        6930	&	$0.152\pm0.036$	&	$-77.7\pm5.3$\\
	7289	&	$0.148\pm0.038$	&	$-78.3\pm3.6$\\
	 \hline
\end{tabular}
\end{table}

We have reduced the \textit{Chandra} data, see Fig.~\ref{a1689_ell}, using the \textit{CIAO} data analysis package  -- version 4.3 -- and the associated calibration database CALDB 4.4.1 \citep{fru+al06}. Out of the 5 exposures available, we have considered the 2 longest ones, Obs ID 6930 (nominal expsoure time of 80 ks) and 7289 (80 ks), both in ACIS-I configuration. The level-1 event files were reprocessed to apply the appropriate gain maps and calibration products and to reduce the ACIS quiescent background. The standard event selection has been applied through the CIAO routine \textit{acis\_process\_events} in ``very faint'' mode configuration. After the exclusions of extended and point sources present in the field and selected with the \textit{vtpdetect} routine, a cleaning of the background light curve has been performed with the \textit{deflare} routine for a total exposure good time of 75.6 ks (Obs ID 6930) and 75.1 (Obs ID 7289).

The images in the 0.7--2 keV band have been corrected for the exposure map,  and then for the presence of the point-sources using the CIAO routine \textit{dmfilth} to produce the image of only extended emission associated to the cluster. A centroid is defined within the source detected with \textit{vtpdetect} and related to the cluster ICM using the CIAO routine \textit{dmstat}. The centroid is then fixed and the routine \textit{dmellipse} is run over the exposure-corrected image of the cluster. 

Measured ellipticity $\epsilon$ and orientation angle are reported in Table~\ref{tab_ellip}. The best-fit results associated to the fit enclosing 80 per cent of the cluster light are considered, see Fig.~\ref{a1689_ell}. Being the two estimates consistent, we fix ellipticity and  angle to the measurements obtained in the exposure 7289. Our estimate of the ellipticity is consistent with the value of $0.16\pm 0.04$ measured in \citet{kaw10} exploiting data from \textit{XMM}.

The surface brightness profiles are then extracted in elliptical annuli. Finally, the two profiles are combined in a single profile by summing the counts and propagating the error on the estimate of the background. We evaluate the extension of the detectable ICM emission by requiring that the signal-to-noise ratio is larger than 2. We measure a maximum radius of $5.9\arcmin$ (1093 kpc). The resulting surface brightness profile is plotted in Fig.~\ref{fig_SB_profile}.

\subsection{X-ray spectroscopic analysis}

The Observation Data Files (ODF) from the XMM-Newton observation have been processed to produce calibrated event files using the XMM-{\em Newton} Science Analysis System ({\tt SAS v10}).

To search for periods of high background flaring, light curves for pattern $\leq 4$ in the 10--12~keV have been produced for two MOS detectors. The soft proton cleaning was performed using a double filtering process similar to the one of \citet{le+mo08}. We extracted a light curve in 100s bins in the 10--12 keV energy band by excluding the central CCD, applied a threshold of 0.20 cts s$^{-1}$, produced a GTI file and generated the filtered event file accordingly. We then extracted a light curve in the 2--5 keV band. The residual intervals of very high background were removed using a 3$\sigma$ clipping algorithm and the light-curves were then visually inspected to remove the background flaring periods not detected by the algorithm. The clean exposure times for the two MOS detectors were $\sim36$~ks. PN data were not used in the analysis because of problems in the background modelling and inconsistencies in the determination of the temperatures with the MOS detectors.

The background in our MOS spectra has been modelled (instead of subtracted) following the procedure developed by \citet{le+mo08}. This procedure is preferable respect to a direct background subtraction because it allows us to use Cash statistics and avoid problems due to the vignetting of the background spectra which would need to be extracted at larger off-axis angle than the cluster spectra where the vignetting of XMM mirrors is not negligible. The main aspects of this procedure are described below, further details can be found on their paper. The background parameter were first estimated in a region free of cluster emission ($10^\prime$--12$^\prime$ annulus). The background model considered included the thermal emission from the Galaxy Halo (HALO, {\em XSPEC model: apec}), the cosmic X-ray background (CXB, {\em XSPEC model: pegpwrlw}), a residual from the filtering of quiescent soft protons (QSP, {\em XSPEC model: bknpower}),  the cosmic ray induced continuum (NXB, {\em XSPEC model: bknpower}) and the fluorescence emission lines ({\em XSPEC model: gaussian}). The Response Matrix File (RMF) of the detector was convolved with all the background components, while the Ancillary Response File (ARF) was convolved only with the first two components (HALO and CXB). The normalization of the QSP component was fixed at the value determined from measuring the surface brightness $SB_\mathrm{in}$ in the 10$^\prime$-12$^\prime$ annulus, and comparing it to the surface brightness $SB_\mathrm{out}$ calculated outside the field of view in the 6-12 keV energy band. Since soft protons are  channeled by the telescope mirrors inside the field of view and the cosmic ray induced background covers the whole detector, the ratio $R_{SB}=SB_\mathrm{in}/SB_\mathrm{out}$ is a good indicator of the intensity of residual soft  protons and was used for the modelling ($Norm(QSP) = 0.03 (1-R_{SB})$). We determined also the 1$\sigma$ error for the background parameters to be used in the fit of the cluster spectra. To fit the cluster spectra, the background parameters (and their 1$\sigma$ errors) were then rescaled by the area where the cluster spectra were extracted. Appropriate correction factors \citep[ dependent on the off-axis angle]{le+mo08} were considered for the HALO and CXB components and a vignetting factor (corresponding to $1.858-0.078r$, with $r$ the distance from the center of the annulus) was applied to the QSP component. All these rescaled values (and their 1$\sigma$ errors) were put in a XSPEC model having the same background components used in the fit of the 10$^\prime$-12$^\prime$ annulus plus a thermal mekal model for the emission of the cluster having the temperature, the abundance and the normalization free to vary. The 1$\sigma$ errors on the parameters were used to fix a range where the normalizations of the background components are allowed to vary. This model was used in addition to the one temperature thermal model used to fit all the spectra and described below.

We divided the emission in five concentric elliptical annuli. The temperature profile is plotted in Fig.~\ref{fig_Te_profile}. Each annulus spectrum was analyzed with XSPEC v12.5.1 \citep{arn96} and fitted with a single-temperature {\tt mekal} model \citep{lie+al95} with Galactic absorption ({\tt wabs} model), in which the ratio between the elements was fixed to the solar value as in \citet{an+gr89}. The fits were performed over the energy range $0.7-8.0$~keV for both of the two MOS detectors. We fitted simultaneously MOS1 and MOS2 spectra to increase the statistics. The free parameters in our spectral fits are temperature, abundance, and normalization. Local absorption is fixed to the Galactic neutral hydrogen column density, as obtained from radio data \citep{kal+al05}, and the redshift to the value measured from optical spectroscopy. 
We used Cash statistics applied to the source plus background\footnote{http://heasarc.gsfc.nasa.gov/docs/xanadu/xspec/manual/XSappendixCash.html}, which requires a minimal grouping of the spectra (at least one count per spectral bin).

\subsection{SZe}

\begin{table}
\centering
\caption{Integrated Compton parameter within 600~kpc, in units of $10^{-10}\mathrm{rad}^2$, as estimated by different observatories. $w_\mathrm{ins}$ is a statistical weight. References are listed in column 4.}
\label{tab_Y_SZ}
\begin{tabular}[c]{lr@{$\,\pm\,$}lcl}
        \hline
        \noalign{\smallskip}
	Observatory	& \multicolumn{2}{c}{$Y_{2500}$} & $w_\mathrm{ins}$	& reference \\ 
        \noalign{\smallskip}
         \hline
         BIMA/OVRO	& 2.0	& 0.4	& 1/2	& \citet{bon+al06}\\
	 BIMA/OVRO	& 2.8	& 0.5	& 1/2	& \citet{ree+al02}\\
	AMIBA		& 2.8	& 1.1	& 1		& \citet{hua+al10}\\
	SuZIE		& 3.6	& 1.2	& 1/2	& \citet{hol+al97}\\
	SuZIE		& 4.4	& 1.2	& 1/2	& \citet{ben+al04}\\
	WMAP		& 2.3	& 0.7	& 1		&  \citet{lie+al06}\\
	SKA			& 2.3	& 0.8	& 1		& \citet{gra+al10}\\
	SCUBA		& 5.8	& 2.4	& 0		& \citet{zem+al07}\\
	 \hline
\end{tabular}
\end{table}

We reviewed available SZe analyses of A1689 in literature, see Table~\ref{tab_Y_SZ}. From published profiles and amplitudes, we estimated the parameter $Y$ in a circular region of outer radius $600~\mathrm{kpc}$. The mean over-density within this radius with respect to the cosmic background density is $\sim 2500$ \citep{bon+al06}. Due to angular resolution and instrument sensitivity, analyses exploiting either BIMA/OVRO, AMIBA, SuZIE, WMAP and SCUBA data usually refer to such radius to determine the SZe amplitude. On the other hand, the integrated Compton parameter from the SKA data  \citep{gra+al10} has been provided within $\sim 500 \arcsec$, i.e., within an overdensity radius of 500, way larger than the X-ray spectroscopic coverage of the cluster. In order to have a coherent combined X-ray plus SZe analyses, we rescaled their result within $600~\mathrm{kpc}$ assuming the profile in \citet{bon+al06}. Relativistic corrections were applied when necessary. 

Since each analysis assumed a fixed parameterization for the density and temperature profile, usually the isothermal $\beta$-model, we added an error of 10 per cent accounting for dependence on modelling \citep{ben+al04,lia+al10}. This systematic error, which was added in quadrature, was usually bested by the statistical uncertainty inferred with standard error propagation, so that its exact value does not matter significantly. Most of the analyses exploited fixed scale radii and slopes for the density profiles from independent X-ray analyses. This approach underestimates parameter degeneracy and statistical uncertainties, so we added in quadrature a further error of 25 per cent. The exception is the analysis in \citet{bon+al06} which fitted the scale radius together with the amplitude to the visibility map. The additional error in this case is not needed.

Some data sets were analysed by independent groups. We considered all published results and weighted each analysis in such a way that the statistical weight of each instrument is the same, see the $w_\mathrm{inst}$ parameter in Table~\ref{tab_Y_SZ}. Finally, measured SZ amplitudes from SCUBA are consistently higher than results inferred from low-frequency measurements \citep{zem+al07}. We did not consider the SCUBA result in our investigation. The final weighted mean for the integrated Compton parameter within 600~kpc is $Y_{2500} =2.5 \pm 0.4$.

\section{The gas distribution}
\label{sec_dist}

\begin{table}
\centering
\caption{Contributions to the total $\chi^2$ value under different hypothesis. We consider: $\eta =0$ and $\delta SB^\mathrm{syst}=0$ (col. 3); $\eta \ge 0$ and $\delta SB^\mathrm{syst}=0$ (col. 4); $\eta =0$ and $\delta SB^\mathrm{syst}=10^{-3}$ (col. 5).}
\label{tab_chi_square}
\begin{tabular}[c]{l|c|ccc}
        \hline
        \noalign{\smallskip}
						&$N_{data}$& $\delta SB^\mathrm{syst}=0$	&	$\eta  \ge 0$&$\delta SB^\mathrm{syst}=10^{-3}$ \\
	\hline
		$\chi^2_{X+SZ}$	&68	&422.7	&422.7	&80.7\\	
		$\chi^2_{SB}$		&56	&415.9	&415.9	&73.9\\
		$\chi^2_{T}$		&5	&4.3		&4.3		&4.2	\\
		$\chi^2_{Y}$		&7	&2.5		&2.5		&2.5 \\
 \hline
\end{tabular}
\end{table}

\begin{table*}
\begin{tabular}[c]{r@{$\,\pm\,$}lr@{$\,\pm\,$}lr@{$\,\pm\,$}lr@{$\,\pm\,$}lr@{$\,\pm\,$}lr@{$\,\pm\,$}lr@{$\,\pm\,$}lr@{$\,\pm\,$}l}
        \hline
        \noalign{\smallskip}
        	\multicolumn{2}{c}{} &	\multicolumn{10}{c}{Electronic density profile} & \multicolumn{4}{c}{Temperature profile}      \\
         \multicolumn{2}{c}{$e_\Delta$} & \multicolumn{2}{c}{$n_0 [\times10^{-4}\mathrm{cm^{-3}}]$} & \multicolumn{2}{c}{$r_\mathrm{P} [\mathrm{kpc}]$}	&\multicolumn{2}{c}{$r_\mathrm{t}/r_\mathrm{c}$}	& \multicolumn{2}{c}{$\beta$}	& \multicolumn{2}{c}{$\gamma$}		& \multicolumn{2}{c}{$T_0 [\mathrm{KeV}]$}	& \multicolumn{2}{c}{$r_\mathrm{T}/r_\mathrm{c}$}    \\
         0.66(.53)	&0.21	&0.31(.28)	&0.05	&47.2(47.1)	&1.0	&13.8(13.9)	&1.5	&0.464(.464)	&0.005	&1.7(1.7)	&0.4	&9.79(9.82)	&0.19	&16.4(16.5)	&1.6	\\
      \hline
\end{tabular}
\caption{Inferred model parameters for the electronic density profile and the temperature profile.}. Central location and dispersion are computed as mean and variance of the PDF. Maximum likelihood values are reported in parentheses.
\label{tab_profiles}
\end{table*}

\begin{figure}
       \resizebox{\hsize}{!}{\includegraphics{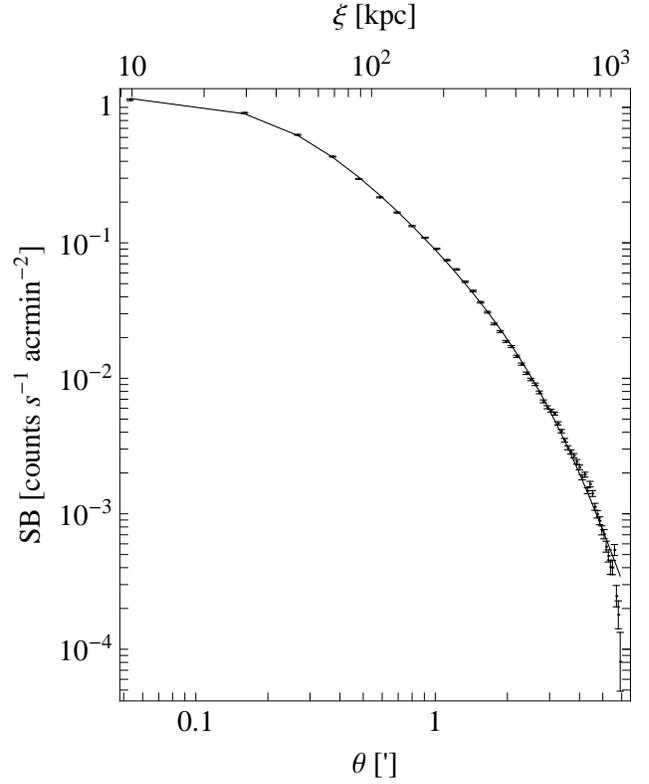}}
       \caption{Radial profile along the projected major axis of the 0.7--2~keV surface brightness measured by Chandra (points with error bars) in elliptical annuli. The full line plots the surface brightness profile predicted by the best fit model.}
	\label{fig_SB_profile}
\end{figure}

\begin{figure}
       \resizebox{\hsize}{!}{\includegraphics{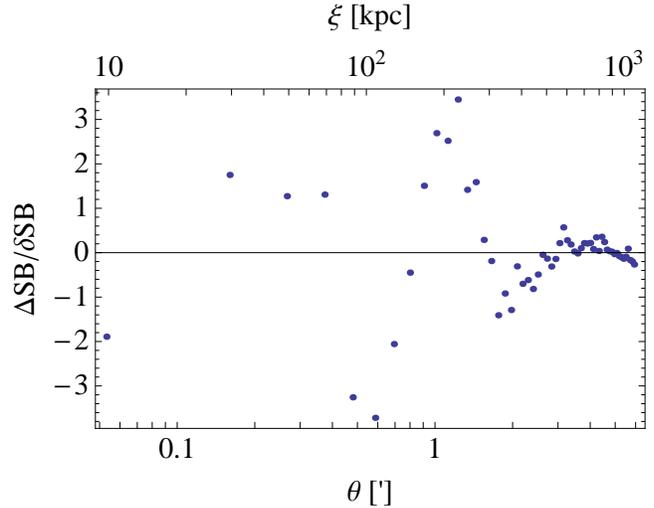}}
       \caption{Radial profile along the projected major axis of the residuals $\Delta SB$ between observed and predicted surface brightnesses. Each scatter is in units of the corresponding observational error $\delta SB$.
       }
	\label{fig_Delta_SB_profile}
\end{figure}

\begin{figure}
       \resizebox{\hsize}{!}{\includegraphics{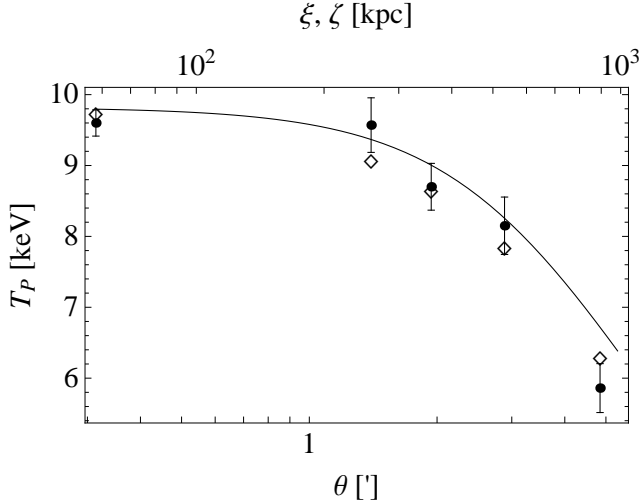}}
       \caption{Radial profile along the projected major axis of the projected temperature measured by XMM (points with error bars) in elliptical annuli. Diamonds denote the spectroscopic-like temperatures predicted by the best fit model. The projected spectroscopic-like temperature is a function of the elliptical radius $\xi$. The full line represents the temperature of the gas distribution $T_\mathrm{e}$ as measured along the ellipsoidal radius $\zeta$.
       }
	\label{fig_Te_profile}
\end{figure}

\begin{figure}
       \resizebox{\hsize}{!}{\includegraphics{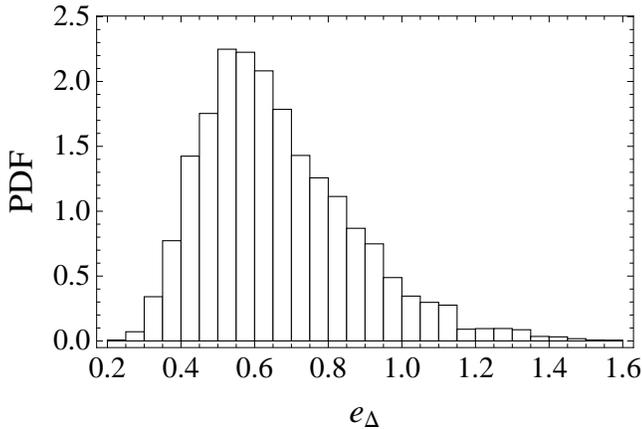}}
       \caption{Posterior probability distribution of the elongation $e_\Delta$.}
	\label{fig_pdf_e_Delta}
\end{figure}

We model the 3D electronic density and temperature with parametric profiles. Distributions are assumed to be coaligned and ellipsoidal, with constant eccentricity and orientation. Profiles are taken from \citet{vik+al06,ett+al09}. For the density profile in the intrinsic coordinate system, we use
\beq
\label{n_prof}
n=n_0 \left( \frac{\zeta}{r_\mathrm{c}} \right)^{-\eta} \left[ 1+\left( \frac{\zeta}{r_\mathrm{c}} \right)^2 \right]^{\eta/2-3\beta/2} \left[ 1+\left( \frac{\zeta}{r_\mathrm{t}} \right)^3 \right]^{-\frac{\gamma}{3}},
\eeq
where $n_\mathrm{0}$ is the central electron density, $r_{\rm c}$ is the core radius, $r_\mathrm{t}(>r_\mathrm{c})$ is the tidal radius, $\beta$ is the slope in the intermediate regions, $\eta$ is the inner slope and  $(0 \le) \gamma (\le 2.5)$ determines the outer slope.  For the temperature profile, we use
\beq
\label{T_prof}
T=T_0 \frac{  (\zeta/r_{T})^{-a_\mathrm{T}} }{  [1+ (\zeta/r_{T})^{b_\mathrm{T}}]^{c_\mathrm{T}/b_\mathrm{T}}  } \frac{    (\zeta/r_\mathrm{cool})^{a_\mathrm{cool}} +T_\mathrm{min}/T_0       }{ (\zeta/r_\mathrm{cool})^{a_\mathrm{cool}} +1},
\eeq
where the radius $r_\mathrm{cool}$ and the slope $a_\mathrm{cool}$ characterise the central cool core with central temperature $T_\mathrm{min}$; $T_0$ is the temperature in the intermediate region; the radius $r_\mathrm{T}$ and the slope $c_\mathrm{T}$ describe a decrement at large radii, with the width of the transition region fixed by $b_\mathrm{T}$.

The metallicity is fixed to the mean observed value.

We are now in position to compare observations with theoretical predictions. Values of X-ray surface brightness, spectroscopic-like temperature and SZ decrement for a given set of parameters can be calculated by first plugging the expressions for $n$, Eq.~(\ref{n_prof}), and $T$, Eq.~(\ref{T_prof}), in the corresponding line of sight integrals, and then by evaluating the intergrals according to the prescription in Eq.~\ref{eq_proj1}. Observations and predictions can then be compared with a $\chi^2$ function,
\begin{eqnarray}
\chi_{X+SZ}^2 & =& \chi^2_{SB}+\chi^2_{T}+\chi^2_{Y} \\
&=& \sum_i^{N_{SB}} \left( \frac{SB_{i}^\mathrm{obs}-SB_{i}^\mathrm{th}}{\delta SB_{i}^\mathrm{obs}} \right)^2 +\sum_i^{N_{T}} \left( \frac{T_{\mathrm{P},i}^\mathrm{obs}-T_{\mathrm{P},i}^\mathrm{th}}{\delta T_{\mathrm{P},i}^\mathrm{obs}} \right)^2  \nonumber \\
& +& \sum_i^{N_{SZ}} w_{\mathrm{ins},i} \left( \frac{Y_i^\mathrm{obs}-Y_i^\mathrm{th}}{\delta Y_i^\mathrm{obs}} \right)^2  . \label{eq_chi1}
\end{eqnarray}
Since we are fitting directly to the projected quantities, instead of the core radius $r_\mathrm{c}$ we determine the projected radius $r_\mathrm{P}$. They are related through Eq.~(\ref{eq:tri6}). The remaining scale lengths are determined in units of $r_\mathrm{c}$. The values of $N_{SB}$, $N_{T}$ and $N_{SZ}$ are reported in Table~\ref{tab_chi_square}.

The models in Eqs.~(\ref{n_prof}) and (\ref{T_prof}) have great functional freedom and can adequately describe almost any type of smooth profile for extended radial ranges. On the other hand, over-parameterization could bring modelling of unrealistic features. The projected temperature was measured in 5 elliptical annuli, whereas the profile in Eq.~(\ref{T_prof}) has 8 free parameters. Firstly, since we did not notice a decline in the inner region, we fixed $T_\mathrm{min}=T_0$. Secondly, degeneracy issues do not allow us to disentangle the effects of either slope $c_\mathrm{T}$ or tidal radius $r_\mathrm{T}$. We then fixed $c_\mathrm{T}=0.9$. Thirdly, the extent of the transition region can not be constrained and we used $b_\mathrm{T}=2$. We end up with only 2 free parameters for the temperature profile ($T_0$ and $r_\mathrm{T}$).

In order to find which parameters in the modelling of the density profile were really needed to reproduce the data, we investigated the $\chi_{X+SZ}^2$ under different assumptions and applied some simple information criteria, such as either BIC or AIC \citep{lid07}. Minima of the $\chi_{X+SZ}^2$ were found through a standard downhill simplex method. Since the value of the minimum did not change for either a free $\eta$ or the case $\eta=0$, see Table~\ref{tab_chi_square}, we found no need for an inner power-law-type cusp and set $\eta=0$. We are then left with 5 free parameters for the density profile. The final set of parameters to be determined is then $n_0$, $r_\mathrm{P}$, $\beta$, $r_\mathrm{t}/r_\mathrm{c}$ and $\gamma$ for the density profile,  $T_0$ and $r_\mathrm{T}/r_\mathrm{c}$ for the projected temperature, plus an overall elongation parameter $e_\Delta$.

Statistical errors on the surface brightness measurements were likely underestimated, see Table~\ref{tab_chi_square}. At the maximum likelihood value, we got $\chi_{X+SZ}^2 \simeq 423$, with $\chi^2_{SB} \simeq 416$ (for 56 points). To check if this affects the results, we considered the effect of a larger error on the surface brightness by adding in quadrature a systematic error of $10^{-3}$ to each measurement in the SB profile. As a consequence, at the maximum likelihood the $\chi^2_{SB} (\simeq 74)$ got a more realistic value whereas the values of $\chi^2_{T}$ and $\chi^2_{Y}$ at the minimum were not affected. The best-fit set of parameters was negligibly affected (with a shift $\ls 0.2$ per cent for $e_\Delta$ and variations $\ls 2$ per cent on other parameters). We can than conclude that a larger error on the SB does not affect our analysis. In fact, best fit values does not change significantly whereas the final estimated error on the elongation, the parameter we are more interested in, is dominated by the statistical uncertainty on the SZe amplitude.

The best fit model, whose model parameters are summarized in the bracketed values in Table~\ref{tab_profiles}, reproduce well the data. Observed values of the surface brightness are plotted in Fig.~\ref{fig_SB_profile} versus the predictions of the best fit model. The scatter plot is in Fig.~\ref{fig_Delta_SB_profile}, where the gap $\Delta SB$ between model prediction and observed value at each radius is represented in units of the observational error $\delta SB$. Most of the values derived with the best fit model are within 1$\sigma$ from the corresponding measured value, and nearly all of them are within 2$\sigma$.

Projected temperature profiles are plotted in Fig.~\ref{fig_Te_profile}. The measured spectroscopic temperatures are very well reproduced by the best fit model. We also plot the gas temperature $T_\mathrm{e}$. Whereas the spectroscopic-like temperature is plotted as a function of the projected elliptical radius $\xi$, the intrinsic temperature $T_\mathrm{e}$ refers to the three-dimensional gas distribution and is a function of the ellipsoidal radius $\zeta$.

The integrated Compton parameter within 600~kpc predicted by the model is $Y \simeq 2.5$, which reproduces perfectly the weighted mean of the observed data. This is expected in our triaxial approach since the free parameter $e_\Delta$ is responsible for the good fit of X-ray and SZ data at the same time.

To asses realistic probability distributions for the parameter we performed a statistical Bayesian analysis. The Bayes theorem states that
\beq
\label{baye}
p(\bfP | \bfd) \propto {\cal L}( \bfP|\bfd) p(\bfP),
\eeq
where $p(\bfP | \bfd)$ is the posterior probability of the parameters $\bfP$ given the data $\bfd$, ${\cal L}( \bfP|\bfd)$ is the likelihood of the data given the model parameters and $p(\bfP)$ is the prior probability distribution for the model parameters. 

As likelihood, we used ${\cal L}_{X+SZ} \propto \exp\left\{ {-\chi_{X+SZ}^2/2}\right\}$. As priors, we used flat probability distributions. We explored the parameter space by running four Markov chains. We checked for chain convergence by verifying that the standard var(chain mean)/mean(chain var) indicator was less than 1.2. Results are summarised in Table~\ref{tab_profiles}. Central location and dispersion for each parameter are the mean and the variance, respectively, of the corresponding marginalised posterior probability function (PDF). Reassuringly, estimates based on the maximum likelihood investigation are fully compatible with the results from the Bayesian analysis.

The posterior probability distribution for the elongation is plotted in Fig.~\ref{fig_pdf_e_Delta}. The distribution is peaked in correspondence of the maximum likelihood value with a long tail at large values, corresponding to rounder configurations. The mean of the distribution, see Table~\ref{tab_profiles}, is then larger than the best-fit value, even if still compatible within $1\sigma$ confidence level.

\section{Additional sources of error}
\label{sec_addi}

Other sources of statistical and systematic uncertainty that affect the measurement of the elongation have to be considered together with the photon-counting statistical uncertainties of the X-ray images and spectra, and the statistical uncertainty of the SZe observations, which determined the statistical error on $e_\Delta$ estimated in the previous section. The elongation enters in the equations as a overall factor of proportionality. Additional sources of error are then similar to what already discussed in detail for the determination of the Hubble parameter \citep{ree+al02,bon+al06,def+al05}. Among the statistical contributions, important roles are played by: the uncertainty in Galactic $N_\mathrm{H}$ ($\sim 1$ per cent), the small-scale clumps in the intra-cluster gas ($\gs 0$ per cent), SZe point sources in the field ($\sim \pm 8$ per cent), kinetic SZ effect  ($\sim \pm 8$ per cent) and CMB anisotropies ($\ls \pm 2$ per cent), contribution from the X-ray background ($\sim \pm 2$ per cent). Finally, recent analyses estimated the error on the Hubble constant to be $\sim \pm 2$ per cent \citep{kom+al11}, which translates in a similar error on $e_\Delta$.

Systematic contributions come from the presence of radio halos and relics ($\sim \pm 3$ per cent), the X-ray absolute flux calibration ($\sim \pm 5$ per cent) and X-ray temperature calibration ($\sim \pm 7.5$ per cent). The absolute calibration of a single instrument for SZe detection brings about an uncertainty of $\ls \pm 8$ per cent. Since we considered several observatories, the corresponding error is reduced to $\ls \pm 8/n_\mathrm{inst}^{1/2}$ per cent. We consider a further error from the peculiar modelling we used. We looked for the minimum of $\chi^2_{X+SZ}$ by modelling only the data points beyond a central cut of 100~kpc with a simple $\beta$-model for the density profile. The shift in the minimum was $\Delta e_\Delta \ls 0.02$.

We added all the discussed independent sources of error in quadrature and ended up with an additional uncertainty of $\delta e_\Delta^\mathrm{sys}=0.11$.

\section{Deprojection}
\label{sec_depr}

\begin{figure*}
\centering
\includegraphics[width=18cm]{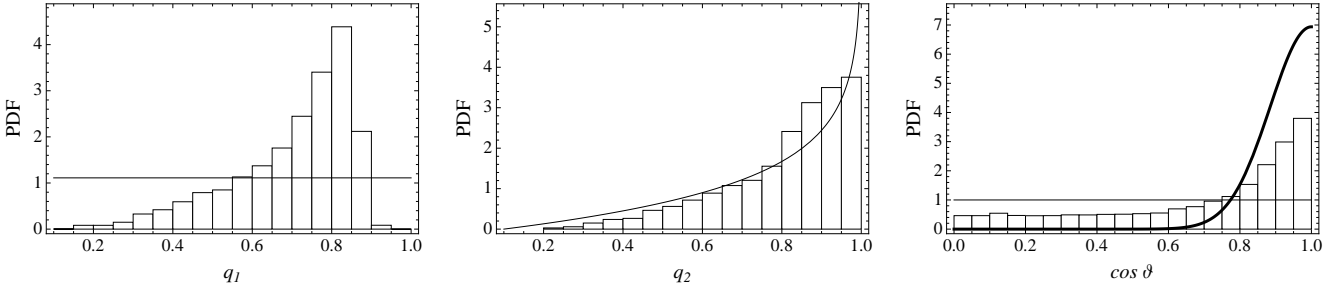} 
\caption{PDFs for the intrinsic parameters under the prior hypotheses of flat $q$-distribution and random orientation angles. Panels from the left to the right are for $q_1$, $q_2$, and $\cos \vartheta$, respectively. The posterior PDFs are plotted as white histograms. Thin lines in the $q$-panels denote the a priori flat $q$-distribution. The thin and the thick line in the $\vartheta$  panel represent a priori random or biased orientation, respectively.}
\label{fig_pdf_All}
\end{figure*}

\begin{table}
\centering
\begin{tabular}[c]{r@{$\,\pm\,$}lr@{$\,\pm\,$}lr@{$\,\pm\,$}l}
        \hline
        \noalign{\smallskip}
        	\multicolumn{2}{c}{$q_1$} &	\multicolumn{2}{c}{$q_2$} & \multicolumn{2}{c}{$\cos \vartheta$}      \\
	\hline
         0.70		&0.15	&0.81	&0.16	&0.70	&0.29	\\
      \hline
\end{tabular}
\caption{ 
Inferred  intrinsic parameters for shape (axis ratios $q_1$ and $q_2$) and orientation (cosine of the inclination angle $\cos \vartheta$). Central values and dispersions are the mean and the standard deviation of the PDF.}
\label{tab_3D_par}
\end{table}

The combined X-ray plus SZ analysis allow us to infer the width of the cluster in the plane of the sky (parameterized in terms of the ellipticity $\epsilon$) and its size along the line of sight (expressed as the elongation $e_\Delta$). We have to use these two observational constraints to infer the intrinsic shape of the cluster ($q_1$ and $q_2$) and its orientation ($\vartheta$ and $\phi$). The problem is clearly under-constrained \citep{ser07}. The use of some a priori hypotheses on the cluster shape can help to disentangle the intrinsic degeneracy. We apply here some Bayesian methods already employed in gravitational lensing analyses \citep{ogu+al05,cor+al09,ser+al10,se+um11}.

As likelihood function we exploited
\begin{eqnarray}
{\cal L}& = & \frac{1}{(2\pi)^{1/2} \sigma_\epsilon} \exp \left\{ -\frac{[\epsilon^\mathrm{obs}-\epsilon]^2}{2\sigma_\epsilon^2}  \right\} \\
& \times & P(e_\Delta-\Delta e_\Delta) \exp \left\{ -\frac{1}{2}\left( \frac{\Delta e_\Delta}{\delta e_\Delta^\mathrm{sys}} \right)^2\right\} . \nonumber
\end{eqnarray}
Ellipticity $\epsilon$ and elongation $e_\Delta$ are functions of $q_1$, $q_2$, $\vartheta$, and $\phi$. $P(e_\Delta)$ is the marginalized posterior probability distribution for the elongation parameter obtained in Sec.~\ref{sec_addi}. The distribution was smoothed using a Gaussian kernel estimator \citep{vio+al94,ryd96}. We parameterized the additional statistical and systematic uncertainty on the elongation as a shift following a normal distribution with dispersion equal to $\delta e_\Delta^\mathrm{sys}$ \citep{dag03}. Since the systematic error is quite smaller than the dispersion the impact on the final results is negligible.

As prior for the intrinsic shape, we considered a flat distribution for the axial ratios in the range $q_\mathrm{min}<q_1 \le 1$ and $q_1 \le q_2 \le 1$. Probabilities are defined such that the marginalized probability $P(q_1)$ and the conditional probability $P(q_2|q_1)$ are constant. The probabilities can then be expressed as
\beq
\label{flat1}
p(q_1) =1/(1-q_\mathrm{min})
\eeq 
for the full range $q_\mathrm{min}<q_1 \le 1$ and
\beq
\label{flat2}
p(q_2|q_1) = (1-q_1)^{-1}
\eeq 
for $q_2 \ge q_1$ and zero otherwise. The resulting probability for $q_2$ is then 
\beq
p(q_2) =\frac{1}{1-q_\mathrm{min}} \ln \left(\frac{1-q_\mathrm{min}}{1-q_2}\right).
\eeq 
A flat distribution allows also for very triaxial clusters ($q_1 \ls q_2 \ll1$), which are preferentially excluded by $N$-body simulations. We fixed $q_\mathrm{min}=0.1$.

For the orientation, we considered a population of randomly oriented clusters with
\beq
\label{flat3}
p(\cos \vartheta) = 1
\eeq 
for $0 \le \cos \vartheta \le 1$ and
\beq
p(\phi)=\frac{1}{\pi}
\eeq
for $-\pi/2 \le \phi \le \pi/2$.

\begin{figure}
	\resizebox{\hsize}{!}{\includegraphics{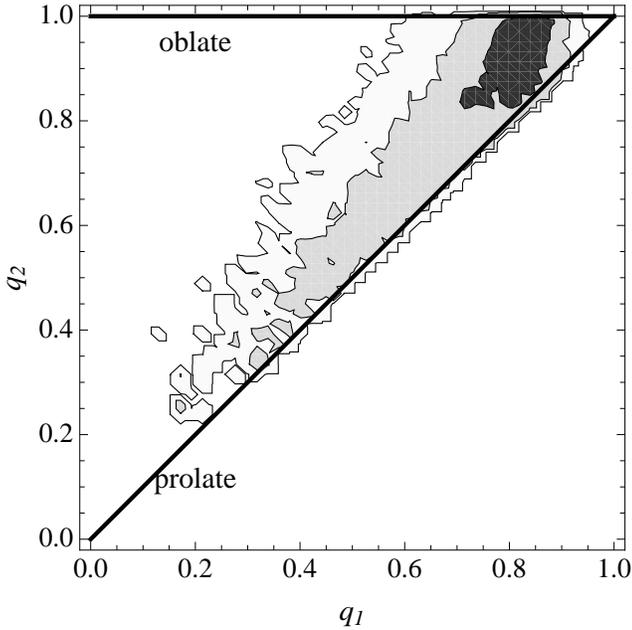}}
	\caption{Contour plot of the marginalised PDF for the the axial ratios $q_1$ and $q_2$. Contours are plotted at fraction values $\exp (-2.3/2)$, $\exp(-6.17/2)$, and $\exp(-11.8/2)$ of the maximum, which denote confidence limit region of 1-, 2- and 3-$\sigma$ in a maximum likelihood framework, respectively. Thick lines denote the loci of points compatible with either prolate or oblate configurations.}
\label{fig_pdf_q1_q2}
\end{figure}

\begin{figure}
	\resizebox{\hsize}{!}{\includegraphics{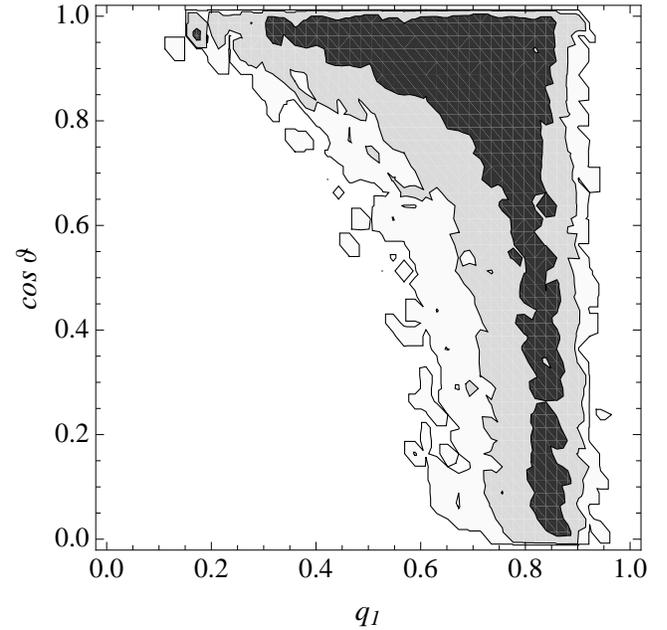}}
	\caption{Contour plot of the marginalised PDF for the the axial ratio $q_1$ and the inclination $\cos \vartheta$. Contours are plotted at fraction values $\exp (-2.3/2)$, $\exp(-6.17/2)$, and $\exp(-11.8/2)$ of the maximum, which denote confidence limit region of 1-, 2- and 3-$\sigma$ in a maximum likelihood framework, respectively.}
\label{fig_pdf_q1_costheta}
\end{figure}

The posterior distribution were investigated by running four Markov chains and checking for convergence. Results are summarized in Table~\ref{tab_3D_par} and Fig.~\ref{fig_pdf_All}. Inferred distributions are reassuringly dominated by the likelihood. The prior on the axial ratios plays an heavy role only for the final distribution of $q_2$, which is not well constrained by data. The distribution of $q_1$ is peaked at $\sim 0.8$ with a tail in correspondence of more triaxial shapes. The spherical hypothesis is ruled out by the very precise measurement of the projected ellipticity. Even if a priori orientations are random, a posteriori biased orientation are favoured. In Fig.~\ref{fig_pdf_All} we compare the inferred distribution for the orientation of the major axis with results from semi-analytical \citep{og+bl09} and numerical \citep{hen+al07} investigations, which showed a large tendency for lensing clusters alike A1689 to be aligned with the line of sight. Such condition can be expressed as \citep{cor+al09}
\beq
\label{nbod5}
p(\cos \vartheta) \propto \exp \left[-\frac{(\cos \vartheta -1)^2}{2 \sigma_\vartheta^2}\right],
\eeq
with $\sigma_\vartheta=0.115$. Noteworthily, the inferred distribution for $\cos \vartheta$ is quite similar to what expected for biased clusters. 

Two dimensional posterior probabilities can provide further insights, see Figs.~\ref{fig_pdf_q1_q2} and~\ref{fig_pdf_q1_costheta}. Mildly triaxial shapes ($0.7 \ls q_1 \ls q_2 \ls 0.9$) are favoured, see Fig.~\ref{fig_pdf_q1_q2}. Oblate configurations ($q_2 =1$) are generally excluded, apart from a small region near $q_1 \simeq 0.8$. On the other hand, prolate shapes ($q_1=q_2$) can reproduce the observational constraints. The major axis is preferentially aligned with the line of sight ($\cos \vartheta =1$), see Fig.~\ref{fig_pdf_q1_costheta}. Orientations in the plane of the sky ($\cos \vartheta = 0$) or intermediate inclinations ($\cos \vartheta \sim 0.5$) are compatible with mildly triaxial shapes ($q_1 \simeq 0.8$). Very triaxial shapes, $q_1 \gs 0.3$, are associated only with aligned configurations ($\cos \vartheta \ls 1$).

\section{Conclusions}
\label{sec_conc}

Knowledge of the intrinsic shape and orientation of halos is crucial to unbiased determinations of their masses and concentrations and to infer their hydrodynamical status. We have discussed a method to determine the geometrical properties of the ICM in rich galaxy clusters. The method is direct and the only hypothesis is that the cluster halo is approximately ellipsoidal. No assumption is needed about hydrostatic equilibrium. We exploited combined information from deep X-ray and SZe observations to constrain the elongation of the cluster gas distribution. This measurement of the size along the line of sight can be used with the measured width in the plane of the sky to infer the intrinsic form. Since the problem is under constrained, Bayesian inference has to be deployed. We applied the method to A1689 and found that the gas distribution is preferentially triaxial and elongated along the line of sight, in agreement with complementary recent results derived with lensing analyses. \citet{se+um11} found a minor to major axis ratio $\sim 0.5 \pm 0.2$ for the shape of the matter distribution. Similar values were found in \citet{cor+al09} and \citet{mor+al11}. 

The technique described in the present paper provides direct estimates on shape and orientation by exploiting very simple assumptions. It does not rely on the hydrostatic equilibrium hypothesis and we could infer the 3D form without any assumption derived from $N$-body simulations. The method is a development of the idea first presented in \citet{def+al05} and \citet{ser+al06}. We implemented some significant improvements. $i)$ The method is still parametric but does not rely anymore on the simple isothermal $\beta$ model. The employed profiles can mimic complex features in either the electronic density or the temperature profile. $ii)$ Instead of the central Compton parameter $y_0$, we considered the more reliable integrated Compton parameter. $iii)$ Even if astronomical deprojection is an under-constrained problem \citep{ser07}, we could infer the 3D structure of the cluster without assuming any specific configuration. Our statistical analysis relied on Bayesian methods which enable to make an inference about a number of variables larger than the number of the observed data. On the other hand in \citet{def+al05} and \citet{ser+al06}, the 3D distribution was assumed to be either triaxial and aligned with the line of sight or prolate or oblate. $iv)$ On top of this, we used more recent data which allowed us to infer the density and temperature profiles up to $r \ls 10^3$~kpc. 

Thanks to the \textit{XMM} spectroscopic data and the \textit{Chandra} derived surface brightness profile, X-ray and SZe analyses probe similar regions. The temperature profile obtained from \textit{Chandra} is consistent in shape with the \textit{XMM}-MOS results, but is systematically higher by about 11 per cent, on average. Similar results are reported in the study by \citet{nev+al10} of the cross-calibration uncertainties in the X-ray measurements of galaxy clusters. Furthermore, thanks to the larger radial extension of  \textit{XMM} observations, we could detect a temperature decrement at larger radii. The smaller temperature and the decrement at large radii account for the gap between the estimate of $e_\Delta \sim 0.7$ in the present paper and the rounder value of $e_\Delta \sim 1$ in \citet{def+al05}.

The application of the method to A1689 strengthens the view of a triaxial cluster elongated along the line of sight. The estimated triaxial structure of A1689 together with the inferred orientation offers a coherent scenario explaining at the same time lensing, X-ray and SZ effect observations. A1689 emerges as a very massive, slightly triaxial halo elongated along the line of sight, with a concentration just above what predicted from numerical simulations and not so far from hydrostatic equilibrium.

A triaxial model of the dark matter halos based on the analysis of $N$-body simulations was proposed by \citet{ji+su02}. Later on, \citet{lee+sut03} considered the relation between the gas distribution and the matter profile in this model under the hypothesis of hydrostatic equilibrium. The model was tested by \citet{kaw10}, who found that the observed distribution of projected axis ratios of 70 clusters imaged by the \textit{XMM-Newton} satellite was consistent with predictions. 

According to the results in \citet{ji+su02} and \citet{lee+sut03}, the intracluster gas distribution in hydrostatic equilibrium under the gravity of triaxial dark matter halo with $q_1^\mathrm{matter}= 0.5 \pm 0.2$, as inferred from lensing \citep{se+um11},  should have $q_1\simeq 0.8\pm0.1$. As expected, the gas distribution is rounder than the total matter density and is in good agreement with our estimate of $q_1 =0.70 \pm 0.15$.

The central mass discrepancy between lensing and X-ray estimates assuming hydrostatic equilibrium might be solved by the orientation bias. Effects of triaxiality might play a larger role than deviations from equilibrium. \citet{pen+al09} showed that masses as estimated with lensing and X-ray analyses might agree if the matter halo of A1689 was oriented along the line of sight with an elongation of $\sim 0.6$. Thanks to our combined X-ray and SZ analysis, we derived a direct estimate of the elongation of the gas distribution of $e_\Delta = 0.7 \pm 0.2$. Being the gas rounder than the mass distribution, our estimate of the elongation is fully consistent with the value required to reconcile the discrepant estimates. On the same page, \citet{mor+al11} relying on theoretical extrapolations of results from $N$-body simulations guessed from their combined X-ray plus lensing analysis that the non thermal contribution to the total pressure in A1689 was about 20 per cent.

Finally, our results reduce the over-concentration problem for A1689. Lensing analyses performed under the hypothesis of spherical symmetry derived an unusually high concentration for the dark matter halo of A1689, well in excess of predictions from $N$-body simulation. Triaxiality was proposed to reconcile observations and theoretical predictions \citep{ogu+al05,cor+al09,se+um11}. The shape and orientation of the gas distribution of A1689 resemble those derived with lensing analyses of the dark matter halo \citep{se+um11}. In particular, we retrieved and confirmed the orientation bias that very likely causes the over-concentration problem. The halo of A1689 seems to be elongated along the line of sight and only slightly over-concentrated but still consistent with theoretical predictions.

\section*{Acknowledgements}
M.S. thanks E. De Filippis for early  discussions and E. Reese for some useful explanations on \citet{ree+al10}. We acknowledge the financial contribution from contracts ASI-INAF I/023/05/0 and I/088/06/0. This research has made use of data obtained from the Chandra Data Archive and the Chandra Source Catalog, and software provided by the Chandra X-ray Center (CXC) in the application packages CIAO, ChIPS, and Sherpa.


\setlength{\bibhang}{2.0em}

\end{document}